\documentclass[aps,prd,twocolumn,showpacs,groupedaddress]{revtex4}

\usepackage{graphicx}
\usepackage{color}
\usepackage{slashed}

\def\apj #1 #2 #3 {#1, ApJ, {\bf #2}, #3}
\def\apjl #1 #2 #3 {#1, ApJ, {\bf #2}, L#3}
\def\apjs #1 #2 #3 {#1, ApJS, {\bf #2}, #3}
\def\aap  #1 #2 #3 {#1, A\&A, {\bf #2}, #3}
\def\mnras #1 #2 #3 {#1, MNRAS, {\bf #2}, #3}
\def\pra #1 #2 #3 {#1, Phys.~Rev.~A., {\bf #2}, #3}
\def\prb #1 #2 #3 {#1, Phys.~Rev.~B., {\bf #2}, #3}
\def\prc #1 #2 #3 {#1, Phys.~Rev.~C., {\bf #2}, #3}
\def\prd #1 #2 #3 {#1, Phys.~Rev.~D., {\bf #2}, #3}
\def\pre #1 #2 #3 {#1, Phys.~Rev.~E., {\bf #2}, #3}
\def\prl #1 #2 #3 {#1, Phys.~Rev.~Lett., {\bf #2}, #3}
\def\plb #1 #2 #3 {#1, Phys.~Lett.~B., {\bf #2}, #3}
\def\science #1 #2 #3 {#1, Science., {\bf #2}, #3}
\def\nature #1 #2 #3 {#1, Nature., {\bf #2}, #3}
\def\nphysa #1 #2 #3 {#1, Nucl.~Phys.~A., {\bf #2}, #3}
\def\nphysb #1 #2 #3 {#1, Nucl.~Phys.~B., {\bf #2}, #3}
\def\nphysbs #1 #2 #3 {#1, Nucl.~Phys.~B.~Suppl., {\bf #2}, #3}

\def\h#1{\hbox{${}^{#1}$H}}

\def\h502{\hbox{$ h^{2}_{50}$}}

\def\fun#1#2{\lower3.6pt\vbox{\baselineskip0pt\lineskip.9pt
  \ialign{$\mathsurround=0pt#1\hfil##\hfil$\crcr#2\crcr\sim\crcr}}}
\begin{document}
\bigskip
\bigskip
%

\title{Possible Evidence for Planck-Scale Resonant Particle Production during 
Inflation from the CMB Power Spectrum
}
\author{ 
G.~J.~Mathews$^{1,2}$, M.~R.~ Gangopadhyay$^1$, 
K.~Ichiki$^{3}$, T.~Kajino$^{2,4}$
}
\address{$^1$Center for Astrophysics,
Department of Physics, University of Notre Dame, Notre Dame, IN 46556 }
\address{
$^2$National Astronomical Observatory, 2-21-1, Osawa, Mitaka, Tokyo
181-8588,
Japan
}
\address{$^3$Department of Physics, Nagoya University, Nagoya 464-8602, Japan
}
\address{$^4$
University of Tokyo, Department of Astronomy, 7-3-1
Hongo, Bunkyo-ku, Tokyo 113-0033, Japan }
\date{\today}
\begin{abstract}
 The power spectrum of the cosmic microwave background from both the {\it Planck} and {\it WMAP} data exhibits a slight dip  for multipoles in the range of
 $l= 10-30$.  We show that such a dip could be the result of the resonant creation of  massive particles that couple
 to the inflaton field.   For our
best-fit models, the epoch of resonant particle creation reenters the
horizon at a wave number of $k_* \sim 0.00011 \pm 0.0004 $  ($h$
Mpc$^{-1}$).  The amplitude and location of this feature corresponds
to the creation of a number of degenerate fermion species of mass $\sim (8-11) /\lambda^{3/2} $ $m_{pl}$ during
inflation where $\lambda \sim (1.0 \pm 0.5) N^{-2/5}$ is the coupling constant between the inflaton field and the
created fermion species, while $N$ is the number of degenerate species.  Although the evidence is of
marginal statistical significance,  this could constitute  new observational hints of unexplored physics beyond the Planck scale.
 \end{abstract}

\maketitle

\section{INTRODUCTION}

The {\it Planck} Satellite \cite{PlanckXIII,PlanckXX} has provided the highest resolution yet available in the  determination
of the power spectrum of the cosmic
microwave background (CMB).
Analysis of this power spectrum 
 provides powerful constraints on the physics of
the very early universe \cite{PlanckXX}. 

The  primordial power spectrum is believed to derive from quantum fluctuations 
generated
during the inflationary epoch \cite{Liddle,cmbinflate}.  
 In this paper we discuss a peculiar feature visible in the observed power spectrum 
near multipoles $\ell = 10-30$.  This is an interesting region in the CMB power spectrum because it corresponds to 
angular scales that are not yet in causal contact, so that the observed power spectrum is close to  the true primordial power spectrum.  

\begin{figure}[htb]
\includegraphics[width=3.5in,clip]{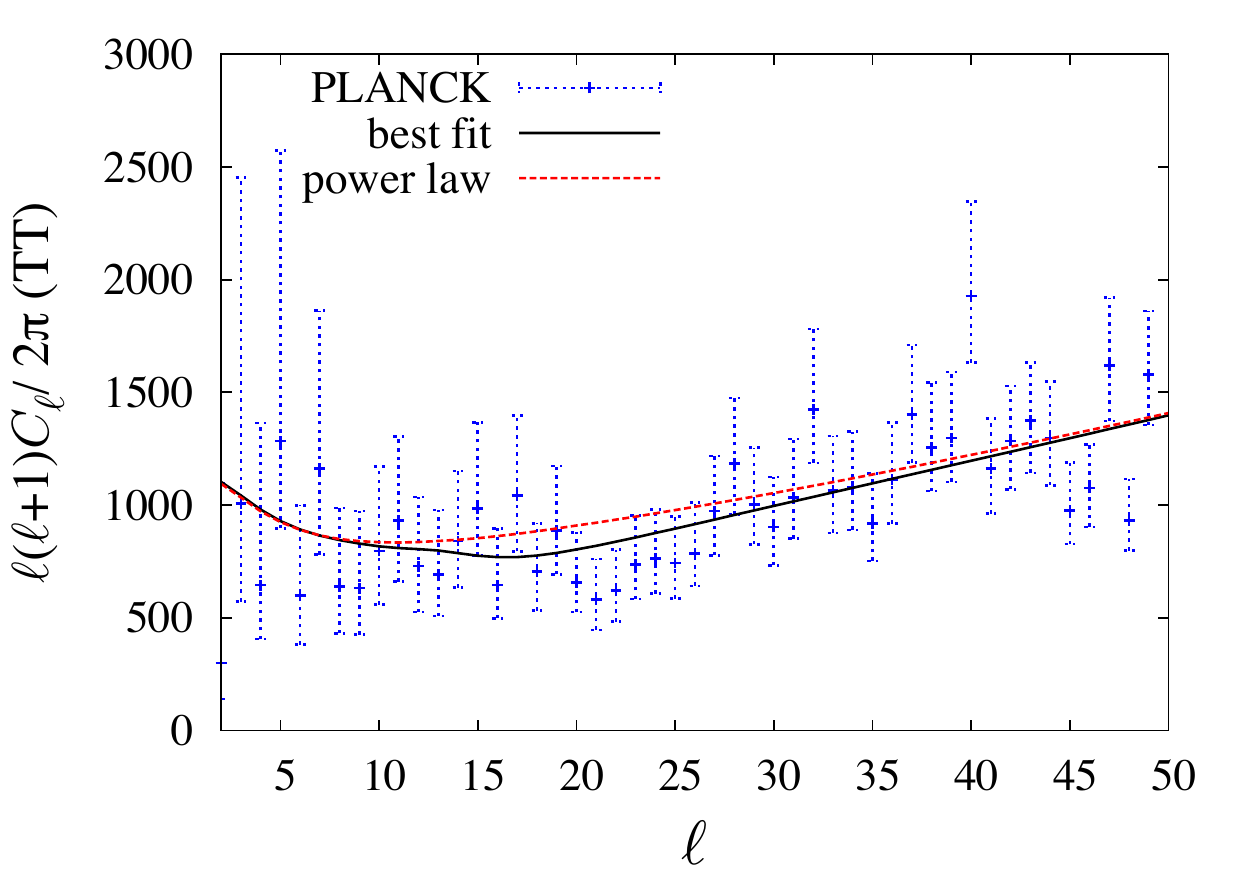} 
\caption{(Color online) CMB power spectrum in the range of $\ell = 3-50$  Points with error bars are from the 
{\it Planck} Data Release \cite{PlanckXIII}.  The dashed line shows the best standard $\Lambda$CDM fit to the {\it Planck} CMB power spectrum
based upon a power-law primordial power spectrum.  The solid line shows the best fit for a model with resonant particle creation during inflation.}
\label{fig:1}
\end{figure}

 An illustration of the {\it Planck} observed power spectrum in this region is shown in Figure \ref{fig:1}.
Although the error bars are large, there is a noticeable systematic deviation  in the range $\ell = 10-30$ below the  best fit based upon the standard $\Lambda$CDM cosmology with a power-law primordial power spectrum.  There is also a well-known possible suppression of the quadrupole moment in the CMB (not shown).
These same  features are  visible in the CMB power spectrum from the Wilkinson Microwave Anisotropy Probe ({\it WMAP}) \cite{WMAP9}, and hence, are likely a true feature in the CMB power spectrum, although it should be noted that in the Planck 
Cosmological Parameters paper \cite{PlanckXX}  
the deviation from a simple power law  in the range $\ell = 10-30$ was deduced to be of  weak  statistical significance due to the large cosmic variance at low $\ell$.   

Nevertheless, a  number of mechanisms have been proposed \cite{Iqbal15} to deal with the suppression of the power spectrum on large scales and low multipoles.  
 In addition to being  an artifact of cosmic variance \cite{PlanckXX,Efstathiou03a}, large-scale power suppression could arise from changes in  the effective inflation-generating potential  \cite{Hazra14}, differing initial conditions at the beginning of inflation 
 \cite{Berera98,Contaldi03,Boyanovsky06,Powell07,Wang08,Broy15,Cicoli14,Das15,Mathews15}, the ISW effect \cite{Das14b}, effects of spatial curvature \cite{Efstathiou03b}, non-trivial topology \cite{Luminet03}, geometry \cite{Campanelli06,Campanelli07}, a violation of statistical anisotropies \cite{Hajian03}, effects of  a cosmological-constant type of dark energy during inflation \cite{Gordon04}, the bounce due to a  contracting phase to inflation \cite{Piao04,Liu13}, the production of primordial micro black-holes  \cite{Scardigli11}, hemispherical anisotropy and non-gaussianity \cite{McDonald14a,McDonald14b}, the scattering of the inflationary trajectory in multiple field inflation by a hidden feature in the isocurvature direction \cite{Wang15}, brane symmetry breaking in string theory \cite{Kitazawa14, Kitazawa15}, quantum entanglement in the M-theory landscape \cite{Holman08}, or loop quantum cosmology \cite{Barrau14}, etc.  
 Most of these works, however, were mainly concerned with the suppression of the lowest moments.  
 
 In the present work, however, we are concerned specifically with suppression of the power spectrum in the range $\ell = 10-30$.  In spite of the weak statistical significance, several recent works \cite{PlanckXX,Hazra14, Kitazawa14, Kitazawa15,Wang15} have deemed it worthwhile to  consider the physical consequences of this deviation, as it could point the way to new interesting physics at and above the Planck scale.
 
 Indeed, in the {\it Planck} cosmological analysis \cite{PlanckXX}   the inflaton potential and the Hubble parameter evolution were reconstructed during the observable part of inflation by using a Taylor expansion of the inflaton potential or $H(\phi)$. When higher-order terms were allowed, both reconstructions found a change in the slope of the potential at the beginning of the observable range, thus better fitting the low-$\ell$ temperature deficit.  As noted in that paper, however,  these models were not significantly favored compared to lower order parameterizations that lead to slow-roll evolution at all times.
 
 Also, in the Planck analysis three distinct methods to reconstruct the primordial power spectrum all  independently found common patterns in the primordial power spectrum of curvature perturbations  related to the dip at $\ell = 10 - 30$ in the temperature power spectrum. 
 
 Although it is of weak statistical significance,  a number of works have proposed explanations of this particular dip anomaly as a possible hint of new physics.  
 One way to explain the anomaly is by a phase transition in the inflation potential \cite{Hazra14}. This is consistent with the abrupt changes in the slope of the inflation potential noted in the Planck reconstruction \cite{PlanckXX}.  
 
 In Ref.~\cite{Hazra14} this feature was fit with a class of models dubbed first order {\it Wiggly Whipped} inflation whereby the field starts rolling from a steeper power law potential and smoothly transitions to a flat power law potential.  This  sharp feature in the inflaton potential produces  a  departure from the initial slow-roll phase, imprinting a large scale suppression in the scalar primordial power spectrum.  The best fits to the dip in  such large field models were  found to have a transition from a faster roll to the slow roll  inflation at an inflaton  field value of  $\phi \approx 15~m_{pl}$. 
 
As noted in that paper, however, in general  this transition and any features in the large field potential produce a suppression of scalar relative to tensor modes at small $k$.  This, however,  is not consistent with the latest Planck results \cite{PlanckXX}  indicating a small tensor to scalar ratio.  This fit  also introduces wiggles in the primordial perturbation.  Such wiggles in  the matter power spectrum might  also be used to constrain this possibility.
 
 In \cite{Kitazawa14,Kitazawa15}  the suppression of low multipoles and the dip for $\ell =10 - 30$ were simultaneously  fit in a string-theory brane symmetry breaking mechanism.  This mechanism splits boson and fermion excitations in string theory, leaving behind an exponential potential that is  too steep for the inflaton to emerge from the initial singularity while descending it. As a result, the scalar field generically "bounces against an exponential wall."  Just as in \cite{Hazra14}, this steep potential then introduces an infrared depression and  a preÐinflationary break in the power spectrum of scalar perturbations, reproducing the observed feature. 
 
 In \cite{Wang15} the dip at $\ell = 10-30$ is explicitly related to the CMB cold spot with an angular radius of $\sim 10^o$ noted in both the Planck \cite{PlanckXIII} and the WMAP \cite{WMAP9} sky maps in the direction $(l,b) = (209^o,-57^o)$.   In their scenario, this could be due to a scattering of  a multiple-field  inflationary trajectory off of a hidden feature in the isocurvature direction.  The inflaton then loses some energy.  If only a patch of the sky hits that feature due to stochastic fluctuations then a cold spot in the sky and a corresponding dip in the temperature power spectrum ensues.

 In the present work, however, rather than to address the implications for the inflation-generating potential, we consider the 
 possibility that  new trans-Planckian physics occurs  near the end of the
inflation epoch corresponding to  the resonant creation \cite{chung00,Mathews04} of Planck-scale particles that
 couple to the inflaton field.  Our best fit is  shown  by the solid line in Figure \ref{fig:1} which  we describe in detail in the following sections.

This interpretation has the
intriguing aspect that, if correct, an opportunity emerges to use the
CMB  to probe  properties of new particle species that existed  at and above the  Planck scale ($m_{pl} \sim
10^{19}$ GeV).  That is the goal of the present work.

Indeed, massive particles generically exist at and above the  Planck
scale due to the compactification schemes of string theory from the Kaluza-Klein states,
winding modes,  string excitations, etc.
Moreover, the coupling of the inflaton to other particle species near the end of inflation is not only natural, but probably required.  
This is because the energy density in the inflaton must be converted to entropy in light or heavy particle species at the end of inflation as a means to reheat the universe.
 Hence, the existence of Planck-scale
mass particles that couple to the inflaton near the end of inflation is a scenario that is both natural and even required.  Moreover, this provides a possible opportunity to uncover new physics in the trans-Planckian regime.  

In our previous study \cite{Mathews04} a similar analysis was made of a possible bump in the CMB in the range of very high multipoles. 
 At that time there appeared to be  an excess power in 
the CMB power spectrum for multipoles in the range $\ell = 2000 - 3500$ in the combined ({\it CBI} \cite{cbi1,cbi2,cbi3}, {\it ACBAR} \cite{acbar},  
{\it BIMA} \cite{BIMA}, and {\it VSA} \cite{VSA}) data, contrary to the expectation from the WMAP results \cite{WMAP1}.  Since that time, however, better high resolution data have eliminated the apparent excess.  

The present analysis, however,  is significantly different from that previous work.  In place of a bump we now seek to fit a dip in the power spectrum. This is achieved by use of a different Lagrangian.  Also, the feature we fit here is at low multipoles and therefore much more likely a part of the primordial spectrum.  Moreover, the deduced particle properties are much different than that of the previous study and even of opposite sign coupling.     Hence, here we present new results on possible resonant fermion particle production during inflation.

\section{Resonant Particle Production during Inflation}

The details of the  resonant particle creation paradigm during inflation have been explained in Refs.~\cite{chung00,Mathews04}.  Indeed, the idea was originally introduced \cite{Kofman94} as a means for reheating after inflation.
Since \cite{chung00} subsequent work \cite{Elgaroy03, Romano08, Barnaby09, Fedderke15}  has elaborated on the basic scheme into a model with coupling between two scalar fields.  
Here,  we summarize essential features of the canonical single fermion field coupled to the inflaton as a means to clarify the possible physics of the $\ell = 10-30$ dip.

In this minimal extension from the basic picture, the
inflaton $\phi$ is postulated to couple to particles
whose mass is of order the inflaton field value.  These particles
are then resonantly produced as the field obtains a critical value
during inflation.  If even a small fraction of the 
inflaton field is affected  in this way, it can produce an observable feature in
the primordial power spectrum. In particular, there can  be either an excess in the power spectrum as noted in \cite{chung00,Mathews04},
or a dip in the power spectrum as described in this paper. Such a dip offers important new clues to the trans-Planckian physics of the early universe.

We note that  particle creation corresponding to an imaginary
part of the effective action of quantum fields has been considered in \cite{Starobinsky02}.  In that case the same creation should occur at the present time.
Thus,  compatibility with
the diffuse $\gamma$-ray background can be used to rule out the possibility of measurable effects from this type of
trans-Planckian particle creation in the CMB anisotropy.
However, the effect of interest here is a perturbation in the simple scalar field due to direct coupling to Planck-mass particles at energies for which the inflation potential is 
comparable to the particle mass and cannot occur at the present time.  The present scenario, therefore is  not
constrained by the diffuse gamma-ray background.

In the simplest slow roll approximation \cite{Liddle,cmbinflate}, the
generation of density perturbations of  amplitude,
$\delta_H(k)$, when  crossing the Hubble
radius is just,
\begin{equation}
\delta_H(k) \approx {H^2
\over 5 \pi \dot \phi}~~,
\label{pert}
\end{equation}
where $H$ is the expansion rate, and $\dot \phi$ is the rate of change of
the inflaton field when the comoving wave number $k$ crosses the
Hubble radius during inflation.  We caution, however, that  resonant particle production
could affect the inflaton field.  In that case the conjugate momentum in
the field $\dot \phi$ could be  altered.  This could cause either an increase or a diminution in
$\delta_H(k)$ (the primordial power spectrum) for those wave numbers which
exit the horizon during the resonant particle production epoch.
In particular,  when $\dot{\phi}$ is accelerated  due to particle production,
it may deviate from the slow-roll condition.  In \cite{chung00}, however, this correction was analyzed
and found to be $<<20 \%$.  Hence, for our purposes we can ignore this correction.

For the application here, we adopt a  positive Yukawa coupling of strength $\lambda$ between the inflaton field $\phi$ and the field $\psi$ of
$N$ fermion species.   This differs from \cite{chung00, Mathews04} who adopted a negative Yukawa coupling.
With our choice, the total Lagrangian density including the inflaton scalar field $\phi$, the Dirac fermion field, and
the Yukawa coupling term is then simply,
\begin{eqnarray}
{\cal L}_{\rm tot} &=& \frac{1}{2}\partial_\mu \phi  \partial^\mu \phi - V(\phi) \nonumber \\
&+& i \bar \psi  \slashed \partial \psi - m \bar \psi  \psi + N \lambda \phi \bar \psi  \psi ~~.
\end{eqnarray}
For this Lagrangian, it is obvious that the fermions have an effective mass of
\begin{equation} 
M(\phi) = m - N \lambda \phi~~.
\end{equation}
This  vanishes
for a critical value of the inflaton field,
$\phi_* = m/N \lambda$.  Resonant fermion production
will then occur in a narrow range of the inflaton field amplitude
around $\phi = \phi_*$.

Note, that the vanishing of the effective mass term with a negative coupling term as in \cite{chung00, Mathews04} requires a positive mass term in the associated free particle Lagrangian.  To achieve this a scenario was adopted in that paper  whereby  the  inflaton $\phi$ 
controls the  fermion mass $\psi$ through the coupling 
\begin{equation}
L_{int}= -[M_f - M_{pl} f(\frac{\phi}{M_{pl}})] \bar{\psi} \psi ~~,
\end{equation}
where $M_f$ is the  fermion mass.
In that case, imposing $f >>1$ leads to a positive mass term and a cancellation of the effective mass is possible.  Here, however, we consider the simpler case of $f << 1$ so that a simple free-particle Lagrangian is sufficient. 

As in \cite{chung00,Mathews04} we label the epoch at which particles are created
by an asterisk.  So, the cosmic scale factor is labeled $a_*$ at the
time $t_*$ at which resonant particle production occurs.  Considering
a small interval around this epoch, one can treat $H = H_*$ as
approximately constant (slow roll inflation).  The number density $n$
of particles can be taken as zero before $t_*$ and afterwards as $n =
n_*[a_*/a(t)]^{3}$.  The fermion vacuum expectation value can then be
written,
\begin{equation}
 \langle \bar \psi \psi \rangle = n_* \Theta (t-t_*) \exp{[-3 H_*(t-t_*)]} ~~.
 \end{equation}
where $\Theta$ is a step function.

Then following the derivation in \cite{chung00,Mathews04}, we have the following modified equation of motion for the scalar field coupled to $\psi$:
\begin{equation}
\ddot \phi + 3 H \dot \phi = -V'(\phi) +  N \lambda \langle \bar \psi \psi \rangle ~~,
\end{equation}
where $V'(\phi) = dV/{d\phi}$.
The solution to this differential equation after particle creation $(t>t_*)$ is then similar to that derived in Refs.~\cite{chung00,Mathews04} but with a sign change for the coupling term, i.e.
\begin{eqnarray}
\dot \phi(t > t_*) &=& \dot \phi_* \exp{[-3H(t-t_*)]}\nonumber \\
& -& \frac{V'(\phi)_*}{3 H_*} \bigl[ 1 - \exp{[-3H(t-t_*)]}\bigr] \nonumber \\
&+&  N \lambda n_* (t-t_*) \exp{[-3 H_*(t-t_*)]} ~~.
\end{eqnarray}
The physical interpretation here is that the rate of change of the scalar field rapidly increases due to the coupling to particles created at the resonance 
$\phi = \phi_*$.  

Then, using Eq.~(\ref{pert})
for the fluctuation as it exits the horizon, and  constant $H \approx H_*$ in the slow-roll condition
along with
\begin{equation}
d\ln{a} = Hdt ~~,
\end{equation}
 one obtains  the perturbation in the primordial power spectrum as it exits the horizon:
\begin{equation}
\delta_H = \frac{[\delta_H(a)]_{N \lambda = 0}}{1 + \Theta (a - a_*)( N \lambda n_*/\vert \dot \phi_*\vert H_*) (a_*/a)^3 \ln{(a/a_*)}} ~~.
\label{deltahnew}
\end{equation}
Here,  it is clear that the power in the fluctuation of the inflaton field will diminish as the particles are resonantly created when the universe
grows to some critical scale factor $a_*$.

Using $k_*/k = a_*/a$, then
the perturbation spectrum Eq.~(\ref{deltahnew})
can be reduced \cite{Mathews04} to a simple 
two-parameter function.
\begin{equation}
\delta_H (k) = \frac{[\delta_H(a)]_{N \lambda = 0}}{1 + \Theta (k-k_*)A (k_*/k)^3 \ln{(k/k_*)}} ~~.
\label{perturb}
\end{equation}
where the amplitude  $A$ and characteristic wave number $k_*$ ($k/k_*
\ge 1$) can be fit to the observed power spectrum from the relation:
\begin{equation}
k_* = \frac{ \ell_* }{ r_{lss}}~~, 
\end{equation}  
where $r_{lss} $ is the
comoving distance to the last scattering surface, taken here to be 14 Gpc.

 The connection between 
resonant particle creation and the CMB temperature fluctuations 
is straightforward.  As usual, temperature fluctuations 
are expanded in spherical harmonics, $ \delta T/T =\sum_l\sum_m
a_{lm}Y_{lm}(\theta,\phi)$ ($2 \le l<\infty$ and $-l \le m \le l$). 
The anisotropies are then described by the angular 
power spectrum, $C_l= \langle |a_{lm}|^2\rangle$, as 
a function of multipole number $l$.  One then merely
requires the conversion from perturbation spectrum $\delta_H (k)$
to angular power spectrum $C_l$.  This is easily 
accomplished using the {\it CAMB} code \cite{Camb}.
When converting to the angular power spectrum,
the amplitude of the narrow particle creation
feature  in $\delta_H(k)$ is spread over many values of $\ell$.
Hence, the particle creation feature looks like a broad dip in the power spectrum.

We have made a  multi-dimensional 
 Markov Chain Monte-Carlo
analysis \cite{Christensen,Lewis} of the 
CMB using the {\it Planck}  data \cite{PlanckXIII} and the {\it CosmoMC} code \cite{Lewis}.   
For simplicity and speed in the present study we
only marginalized over 
parameters which do not alter the matter or CMB transfer functions. Hence, we only varied $A$ and $k_*$, along with the six parameters, 
 $\Omega_b h^2, \Omega_c h^2, \theta, \tau, n_s, A_s$.  Here,  $\Omega_b $ is the baryon content, $ \Omega_c $ is the cold dark matter content,  $\theta$ is the acoustic peak angular scale, $\tau$ is the optical depth, $n_s$ is the power-law spectral index, and $ A_s$ is the normalization.
As usual,
both $n_s$ and $A_s$ are normalized 
at $k = 0.05$ Mpc$^{-1}$. 

\begin{figure}[htb]
\includegraphics[width=3.5in,clip]{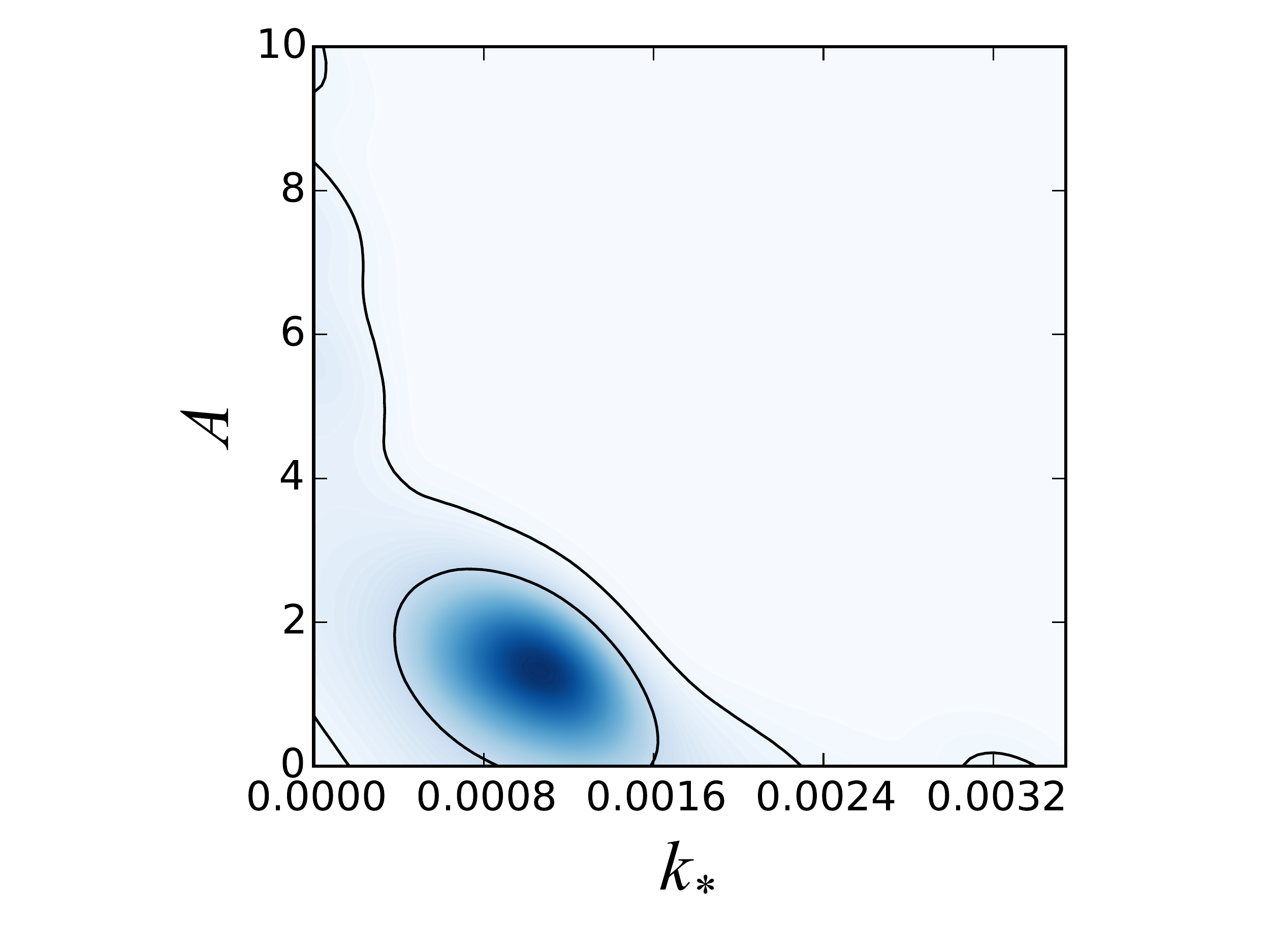} 
\caption{(Color online) Constraints on parameters $A$ and $k_*$ from the
MCMC analysis of the CMB power spectrum.
Contours show 1 and  2$ \sigma$ limits.  The horizontal
axis is in units of ($h$ Mpc$^{-1}$). }
\label{fig:2}
\end{figure}

Figure \ref{fig:2} shows contours of likelihood for the resonant particle creation parameters,
$A$ and $k_*$.  Adding this perturbation to the primordial power spectrum improves the total 
$\chi^2$ for the fit from 9803 to 9798.  One expects that the effect of interest here would only make a small change ($\Delta \chi^2 = 5$) in the overall fit because  
it only affects a limited range of $l$ values with large error bars.  Nevertheless, from
 the likelihood contours we can deduce a mean value of $A = 1.7 \pm 1.5$ with a maximum likelihood value of   $A = 1.5$,  and a mean value of $k_* = 0.0011\pm 0.0004 ~h~{\rm Mpc}^{-1}$

Of course, it is obvious that adding extra parameters should improve the goodness of fit.  One should quantify the statistical significance of the improvement over a simple power-law primordial power spectrum.  A $\Delta \chi^2 = 5$ in the fit corresponds to a 92\% confidence level for two free parameters, hence less than a 2$\sigma$ confidence limit.  To be more precise, the Bayesian information criterion (BIC) can be used to select whether one model is better than another by introducing a penalty term for the number of parameters in the model fit.
Under the assumption that the model errors are independent and obey a normal distribution, then the BIC can be rewritten in terms of $\Delta \chi^2$ as 
BIC$ \approx \Delta \chi^2  + df \cdot \ln{n}$ where $df$ is the number of degrees of freedom in the test and $n$ is the number of points in the observed data. For the 30 multipoles in the range of the fit,  the introduction of 2 new free parameters then corresponds to a BIC$ = 1.8$. Generally, BIC$> 2$ is considered positive evidence for an improvement in the fit.  Hence, one must conclude that the evidence for this fit is statistically weak.  Nevertheless, it is worthwhile to examine the possible physical meaning of the deduced parameters.

\section{Physical Parameters}

The values of $A$ and $k_*$ determined from from the CMB power spectrum
relate to the inflaton coupling $\lambda$ and fermion mass $m$, for a
given inflation model via Eqs.~(\ref{deltahnew}) and (\ref{perturb}). 
\begin{equation}
A  = |\dot{\phi}_*|^{-1} N \lambda
n_* H_*^{-1} ~~.
\end{equation}

The coefficient $A$
can be related directly to the coupling constant $\lambda$ 
using the approximation
\cite{chung00,birrellanddavies,Kofman:1997yn,Chung:1998bt} for the
particle production Bogoliubov coefficient 
\begin{equation}
|\beta_k|^2 = \exp\left( \frac{-\pi k^2}{a_*^2 N \lambda |\dot \phi_*| }\right).
\end{equation}

Then,
\begin{equation}
\label{eq:nstar}
n_* = \frac{2}{\pi^2}\int_0^\infty dk_p \, k_p^2 \, 
|\beta_k|^2 =
\frac{N \lambda^{3/2}}{2\pi^3}
|\dot{\phi}_*|^{3/2}~~ .
\end{equation}
This give us
\begin{eqnarray}
A & = & \frac{ N \lambda^{5/2}}{2 \pi^3}
\frac{\sqrt{|\dot{\phi}_*|}}{H_*}\\
& \approx & \frac{N \lambda^{5/2}}{2 \sqrt{5} \pi^{7/2}}
\frac{1}{\sqrt{\delta_H(k_*)|_{\lambda=0}}}~~.
\label{eq:alamrelation}
\end{eqnarray}
where we have used the usual approximation for the primordial slow
roll inflationary spectrum \cite{Liddle,cmbinflate}.  This means that
regardless of the exact nature of the inflationary scenario, for any
fixed inflationary spectrum $\delta_H(k)|_{\lambda=0}$ without the
back reaction, we have the particle production giving  a dip of the
form Eq.~(\ref{perturb}) with the parameter $A$ expressed in terms of
the coupling constant through Eq.~(\ref{eq:alamrelation}).  Given that
the CMB normalization requires $\delta_H(k)|_{\lambda=0}\sim
10^{-5}$, we then have 
\begin{equation}
A \sim 1.3 N  \lambda^{5/2}.
\end{equation}
Hence, for the maximum likelihood value of $A \sim 1.5 $, we have 
\begin{equation}
\lambda  \approx \frac{(1.0 \pm 0.5)}{N^{2/5}} ~~.  
\label{Nconst}
\end{equation}
 So,   $\lambda \le 1$ requires $N>1$ as expected.

The fermion particle mass $m$ can then be deduced from $m = N \lambda \phi_*$.  From Eq.~(\ref{Nconst}) then we have
$m \approx   \phi_*/\lambda^{3/2}$.
For this purpose, however, one must adopt a specific form for the inflaton potential to determine $\phi_*$ appropriate to the scale $k_*$.
Here, we adopt a general monomial potential whereby:
\begin{equation}
V(\phi) = \Lambda_\phi m_{pl}^4 \biggl(\frac{\phi}{m_{pl}}\biggr)^\alpha~~,
\label{Vphi}
\end{equation}
for which  there is a simple  analytic relation \cite{Liddle} 
 between  the value of $\phi_*$ and the number of e-folds ${\cal N}(k_*)$ between when $k_*$ exits the horizon and the end of inflation, i.e.
\begin{equation}
{\cal N}(k_*) = \frac{1}{m_{\rm pl}^2} \int_{\phi_{end}}^{\phi_*} \frac{V(\phi)}{V'(\phi)} d\phi~~,
\end{equation}
implies
\begin{equation} 
\phi_* = \sqrt{2 \alpha {\cal N}} m_{pl}~~.
\end{equation}

For $k_* = 0.0011\pm 0.0004 ~h~{\rm Mpc}^{-1}$, and $k_H = a_0 H_0 = (h/3000)$ Mpc$^{-1} \sim 0.0002,$ we have ${\cal N} - {\cal N}_*  = ln{(k_H/k_* )} <1$.  Typically one expects  ${\cal N}(k_*) \sim  {\cal N} \sim 50 - 60$.  We note, however, that one can have the number of e-folds as low as ${\cal N} \sim 25$ in the case of thermal inflation \cite{Liddle}.  For standard inflation a monomial potential with   $\alpha = 2$ would have $\phi_* \sim (14- 15)~ m_{pl}$.  However, the limits on the tensor to scalar ration from the {\it Planck} analysis \cite{PlanckXX} rule out $\alpha = 2$ at the 95\% confidence level.  Monomial potentials are more consistent with  $\alpha = 1$ ($\phi_* = (10-11) ~m_{pl}$), or even $\alpha = 2/3$ ($\phi_* = (8-9) ~m_{pl}$).  
Hence,  we have roughly the constraint,  
\begin{equation}
m \sim  (8-11)  ~\frac{m_{\rm pl}}{\lambda^{3/2}}~~.
\end{equation}

\begin{figure}[htb]
\includegraphics[width=3.5in,clip]{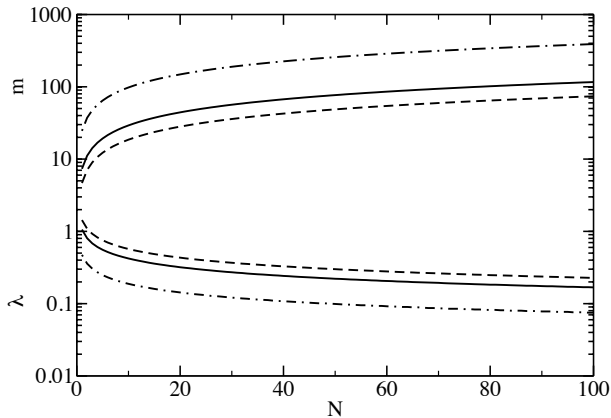}
 \caption{ Implied values of $\lambda$ and $m$ as a function of the number of degenerate species $N$ for a $\phi^{2/3}$ inflaton effective potential experiencing 50 $e$-folds of inflation. Upper curves show the allowed mass.  Lower curves are the coupling constant.  Solid line is for the best-fit normalization $A = 1.5$.  The dashed and dot-dashed lines show the uncertainties due to the upper and lower limits on $A$, respectively.}
\label{fig:3}
\end{figure}

So, one can deduce a family of possible  properties of the resonantly produced particle (i.e. its mass and coupling strength)  in terms of a single parameter, the degeneracy $N$.
This is illustrated in Figure \ref{fig:3} that shows allowed values and uncertainty in the coupling constant and particle mass as a function of the number of degenerate species for a $\phi^{2/3}$ inflaton effective potential experiencing 50 $e$-folds of inflation.

Indeed, it is natural \cite{chung00} to have a large  degeneracy (e.g. $N\sim 10-100$  in Figure \ref{fig:3})  among  trans-Planckian massive particles.  Supergravity and super-string theories generally  contain a spectrum of particles with masses well in excess of the Planck mass.  Moreover,  the compact extra-dimensions
lead to  a tower of nearly degenerate Kaluza-Klein (KK) states \cite{Kolb84,Lewis03}, and as noted above, reheating may require that  some of these particles  couple to the inflaton field near the end of inflation.

\section{Matter Power Spectrum }

It is perhaps obvious that the matter power spectrum will be unaffected by the $\ell = 10-30$ anomaly since, as noted above, this range of multipoles is for the most part not yet in causal contact.  Nevertheless, the largest multipoles affected by this dip are near the scale of the horizon at decoupling.  Hence, as in other studies \cite{Hazra14} for completeness, we examine the impact of this anomaly on the matter power spectrum.   
 
 It is straight forward to determine the matter power spectrum.  To
convert the amplitude of the perturbation as each wave number $k$ enters
the horizon, $\delta_H(k)$, to the present-day power spectrum, $P(k)$,
which describes the amplitude of the fluctuation at a fixed time, one
 utilizes the  transfer function, $T(k)$ \cite{efstathiou} which
is easily computed using the {\it CAMB} code  \cite{Camb} for various
sets of cosmological parameters (e.g.~$\Omega$, $H_0$, $\Lambda$,
$\Omega_B$).  An adequate approximate expression for the structure power
spectrum is then
\begin{equation}
\frac{k^3}{2\pi^2}P(k) = \left( \frac{k}{aH_0} \right)^4 T^2(k)
\delta^2_H(k) \ .
\end{equation}
This expression is only valid in the linear regime,
which in comoving wave number is up to approximately $k ^<_\sim
0.2~h$ Mpc$^{-1}$ and therefore
adequate for our purposes.
However, we also correct for
the nonlinear evolution of the power spectrum \cite{Peacock}.

\begin{figure}[htb]
\includegraphics[width=3.5in,clip]{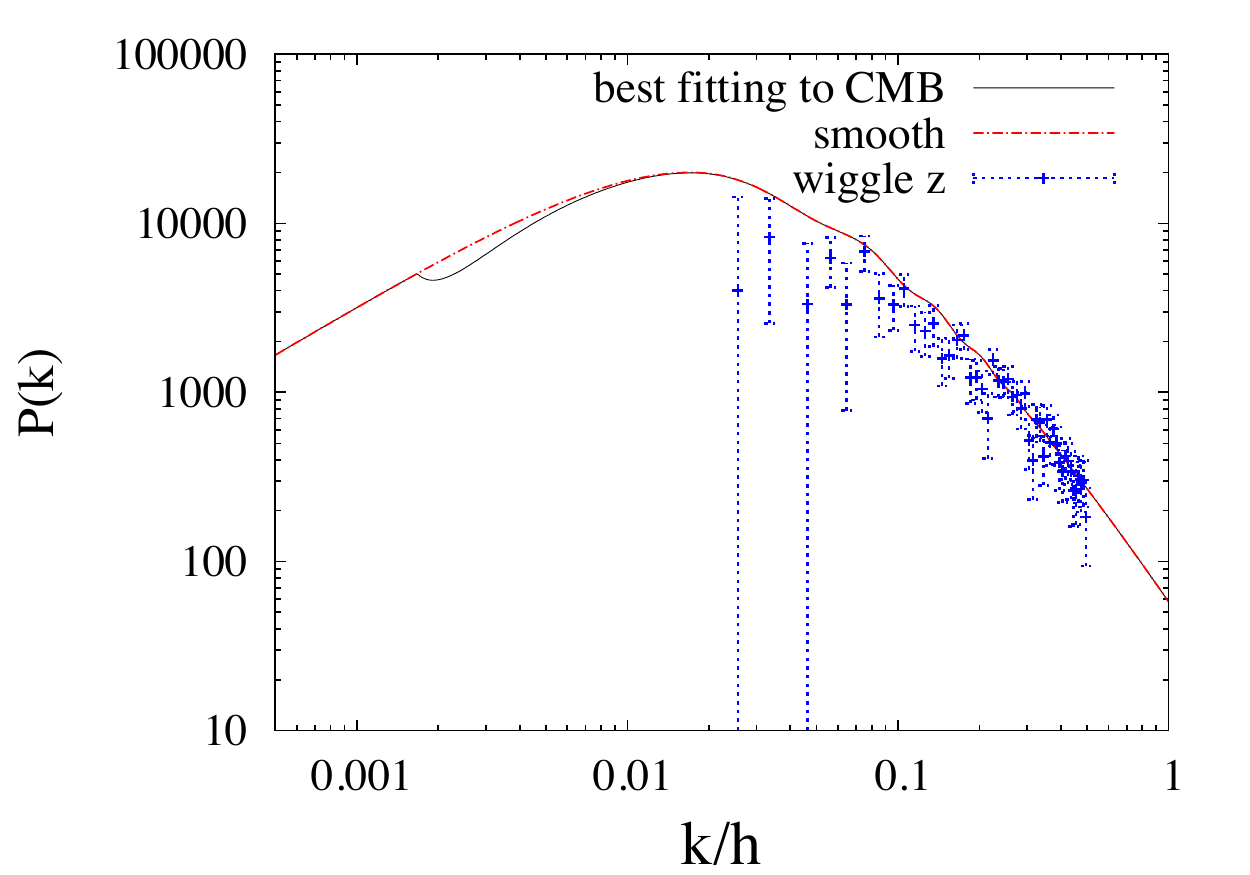}
 \caption{(Color online) Comparison of the observed galaxy cluster function
from  \cite{wiggle}
 with the spectrum implied from
the fits to the matter power spectrum with (solid line)  and without (dashed line)
resonant particle creation during inflation as described in the text.}
\label{fig:4}
\end{figure}

Figure \ref{fig:4} shows the  matter power spectrum from the {\it Wiggle-Z Dark Energy Survey} \cite{wiggle} compared to the computed
maximum likelihood power spectrum with and without
the perturbation due to the resonant particle creation.
Unfortunately, the  perturbation is on a scale too large to be probed by the observed matter power spectrum.

\section{Conclusion}
We have analyzed the $\ell = 10-30$ dip in the {\it Planck} CMB power spectrum in the context of a
model for the creation $N$ nearly degenerate  trans-Planckian massive fermions during inflation.  The  best fit to the  CMB  power spectrum
implies an optimum feature at $k_* = 0.0011 \pm 0.0004~h$\,Mpc$^{-1}$ 
and $A \approx  1.7 \pm 1.5$.  For monomial inflation potentials consistent with the {\it Planck} tensor-to-scalar ratio,  this feature  would correspond to the resonant creation
of nearly degenerate  particles with $m \sim 8-11$ $m_{pl}/\lambda^{3/2}$ and a Yukawa coupling
constant $\lambda$ between the fermion species and the inflaton field of $\lambda
\approx (1.0 \pm 0.5)N^{-2/5}$ for $N$  degenerate fermion species.

Obviously there is a need for more precise determinations of the
CMB  power spectrum for multipoles in the range of $\ell = 10-30$, although this may ultimately be limited by the cosmic variance.

Nevertheless, in spite of these caveats, we conclude that if the present
analysis is correct, this may be one of the first  hints at observational
evidence of new particle physics at the Planck scale.  Indeed, one
expects a plethora of particles at the Planck scale, particularly in
the context of string theory.  Perhaps, the presently observed CMB power
spectrum contains the first suggestion that a subset of such particles may have 
 coupled to the inflaton field leaving a relic signature of their existence in the
CMB primordial power spectrum.

\acknowledgements
Work at the University of Notre Dame is supported
by the U.S. Department of Energy under 
Nuclear Theory Grant DE-FG02-95-ER40934.
Work at NAOJ was supported in part by Grants-in-Aid for Scientific Research of JSPS (26105517, 24340060).
Work at Nagoya University supported by JSPS research grant number 24340048.

\end{document}